\documentclass[a4paper,11pt]{article}
\pdfoutput=1
\usepackage{jheppub} 
\usepackage[T1]{fontenc}

\usepackage[utf8]{inputenc}
\usepackage{amsmath,mathtools}

\usepackage{tensor}
\usepackage{amssymb}
\usepackage{graphicx}
\usepackage{physics}
\usepackage{lipsum}  
\usepackage{bigints}
\usepackage{verbatim}

\usepackage{braket}
\usepackage{cancel}
\usepackage{color}
\usepackage{hyperref}

\newcommand{\be}{\begin{equation}}
\newcommand{\ee}{\end{equation}}
\newcommand{\bea}{\begin{eqnarray}}
\newcommand{\eea}{\end{eqnarray}}
\newcommand{\cphi}{\varphi}

\newcommand{\non}{\nonumber}
\newcommand{\cf}{\mathfrak{f}}

\title{\boldmath On the stability of scale-invariant black holes}

\author[a,b]{S. Boudet,}
\author[a,b]{M. Rinaldi}
\author[a,b]{and S. Silveravalle}

\affiliation[a]{Dipartimento di Fisica, Universit\`{a} di Trento,\\Via Sommarive 14, I-38123 Povo (TN), Italy}
\affiliation[b]{Trento Institute for Fundamental Physics and Applications TIFPA-INFN,\\Via Sommarive 14, I-38123 Povo (TN), Italy}

\emailAdd{simon.boudet@unitn.it}
\emailAdd{massimiliano.rinaldi@unitn.it}
\emailAdd{samuele.silveravalle@unitn.it}

\abstract{Quadratic scale-invariant gravity non minimally coupled to a scalar field provides a competitive model for inflation, characterized by the transition from an unstable to a stable fixed point, both characterized by constant scalar field configurations. We provide a complementary analysis of the same model in the static, spherically symmetric setting, obtaining two Schwarzschild-de Sitter solutions, which corresponds to the two fixed points existing in the cosmological scenario. The stability of such solutions is thoroughly investigated from two different perspectives. First, we study the system at the classical level by the analysis of linear perturbations. In particular, we provide both analytical and numerical results for the late-time behavior of the perturbations, proving the stable and unstable character of the two solutions. Then we perform a semi-classical, non-linear analysis based on the Euclidean path integral formulation. By studying the difference between the Euclidean on-shell actions evaluated on both solutions, we prove that the unstable one has a meta-stable character and is spontaneously decaying into the stable fixed point which is always favoured.}

\begin{document} 
\maketitle
\flushbottom

\section{Introduction}\label{sec: intro}
Quadratic gravity is a well-known minimal extension of General Relativity (GR), which adds, to the usual Einstein-Hilbert term $R$ in the action, quadratic terms obtained by contracting the Riemann tensor $R_{\alpha\beta\mu\nu}$. The classical and quantum properties of this theory were thoroughly investigated already in the late 70's \cite{Stelle:1976gc,Stelle:1977ry}, initially in the context of compact and spherically symmetric objects (see also \cite{Bonanno:2022ibv, Bonanno:2019rsq,Bonanno:2021zoy,Silveravalle:2022lid} for further developments). 

The simplest quadratic Lagrangian, often called Starobinski model, has the form ${\cal L}\propto \sqrt{-g}( R+R^2/m^2)$ \cite{starobinsky_new_1980} and, when used to describe the inflationary expansion of the Universe, turns out to be among the most accurate models when compared to observations \cite{Planck:2018vyg}. The linear and quadratic terms can also be interpreted as the truncation of a Taylor expansion of an analytic function $f(R)$ around flat space. The cosmological and astrophysical implications of $f(R)$ gravity have attracted a flurry activity in the past decades, see e.g. \cite{2010LRR133D,Sotiriou:2008rp} for reviews.

All quadratic models of gravity share the property that, when the curvature is large with respect to a suitable scale, the theory becomes globally scale-invariant, since the coefficients of the quadratic terms in the action are dimensionless. Typically, in these scale-invariant models of gravity, scales are generated by quantum corrections \cite{Cooper:1981byv,RevModPhys.54.729,PhysRevD.7.1888,Salvio:2014soa,Einhorn:2014gfa,Einhorn:2015lzy,Einhorn:2016mws,Edery:2015wha,Salvio:2020axm}. As for the Starobinski model, fundamental scale-invariance naturally leads to flat inflationary potentials  \cite{Khoze:2013uia,Kannike:2014mia,Rinaldi:2014gha,Salvio:2014soa,Kannike:2015apa,Barrie:2016rnv,Tambalo:2016eqr}. 
Finally, black holes in the simplest model of quadratic gravity (${\cal L}=\sqrt{g}R^2$) were studied concerning their stability in \cite{Dioguardi:2020nxr} and the thermodynamics in \cite{Cognola:2015uva,Cognola:2015wqa}.

As mentioned above, the main mechanism to generate scales in a scale-invariant model is assumed to be loop quantum corrections. However, mass scales can emerge naturally also in a classical way. This was first noticed, in the context of inflationary cosmology, in \cite{Rinaldi:2015uvu}. Here, it was shown that the simplest extension of scale-invariant quadratic gravity, obtained by adding a scalar field non-minimally coupled to gravity and with a quartic potential, describes a system that dynamically evolves from an unstable fixed point to a stable one. In both fixed points, the geometry of the two manifolds is that of a flat de Sitter space with two different values for the cosmological constants and the scalar field. In particular, at the stable fixed point the scalar field stabilizes and the coefficient of the linear term in $R$ of the action can be identified as the Planck mass squared, while the others become negligible. During the transition between the two fixed points the Universe expands exponentially, giving rise to a competitive model of inflation \cite{Tambalo:2016eqr,Ghoshal:2022qxk}, which is also robust against quantum corrections \cite{Vicentini:2019etr}.

The aim of the present paper is to explore the possibility that the fixed points mentioned above exist also in the space of spherically symmetric and static solutions. In fact, it is quite easy to show that the equations of motion endowed with the spherically symmetric and static symmetry, yield two asymptotically de Sitter black hole solutions, with arbitrary cosmological constant. Thus, the same theory yields to static solutions and we would like to understand their stability and, eventually, how the transition from the unstable to the stable one happens.

We will show that, indeed, one solution is unstable while the other is stable by following two complementary paths, based on different notions of stability.

On one hand, from a completely classical perspective, the dynamical stability of an exact solution of the field equations can be deduced from the behaviour of linear perturbations. In GR, when we deal with static, spherically symmetric backgrounds, this amounts to linearize Einstein's equations on the given background solution and adopt a harmonic decomposition for the metric perturbation, resulting in the well-known Regge-Wheeler and Zerilli equations \cite{Maggiore:2018sht}. For a scalar-tensor theory instead, one has to consider the dynamics of the scalar field perturbation as well, which, as we will see, turns out to be the only one modified with respect to the GR case. We will then investigate the stability by analyzing the late-time behaviour of the perturbations, which can either diverge, yielding unstable solutions, or go asymptotically to zero. In the second case, the black hole is stable against perturbations, eventually relaxing to a new static configuration.    

On the other hand, we will study the semiclassical properties, by means of the path integral, of the system and determine a criterion to establish which of the solutions is favoured. The task is not easy since it is notoriously difficult to interpret an asymptotically de Sitter black hole in terms of a thermodynamic system in equilibrium, the reason being that the two horizons have different temperatures. Therefore, we need to resort to different considerations. While this approach is less rigorous in assessing the stability of the solutions, it can give relevant insight for the transition between one solution and the other, and on the role of black holes in the cosmological evolution.

The paper is organized as follows: in Sec.\ \ref{sec: summary} we recall the main features of the scale-invariant cosmological solution proposed in \cite{Rinaldi:2015uvu}. We examine the equations of motion in the Einstein frame and in a convenient representation. We also consider an extension of the no-hair theorem in this context. In Sec.\ \ref{sec: solutions} we briefly recall the cosmological solutions and the fixed point structure of the phase space. We then show that the most general scale-invariant tensor-scalar action admits asymptotically (anti) de Sitter black holes with arbitrary mass and arbitrary cosmological constant, together with a possibly non trivial scalar field value. In Sec.\ \ref{sec: thermo stability} we study the stability of the solutions in terms of dynamical classical perturbations and, in Sec. \ref{sec: dyn stability}, in terms of the Euclidean action. We conclude in Sec.\ \ref{sec: conclusions} with some conclusions and remarks.

\section{Quadratic scale-invariant gravity}\label{sec: summary}

We consider the general quadratic scale-invariant action studied in \cite{Rinaldi:2015uvu,Ghoshal:2022qxk}, that is
\begin{equation}\label{cosmoaction}
    \mathcal{I} = \int\mathrm{d}^4x\sqrt{-g}\left[ \frac{\alpha}{36}R^2+\frac{\xi}{6}\phi^2 R-\frac{1}{2}\partial^\mu\phi\partial_\mu\phi-\frac{\lambda}{4}\phi^4  \right],
\end{equation}
where $\alpha,\,\xi$, and $\lambda$ are dimensionless arbitrary positive constants. Strictly speaking (\ref{cosmoaction}) is not the most general quadratic scale-invariant action with one scalar field, having no terms with quadratic combinations of the Ricci and Riemann tensors. However, these terms are known to bring instabilities at quantum level \cite{Stelle:1976gc}, and greatly complicate the study of perturbations, having no simple description in the Einstein frame \cite{Hindawi:1995an}. Some considerations on the general theory will be made in Appendix \ref{sec: appendix a}.\\ 
The metric field equations of (\ref{cosmoaction}) read
\begin{align}\label{metric equation}
&\left(\frac{\alpha}{18}R+\frac{\xi}{6}\phi^2\right)G_{\mu\nu}-\frac{1}{2}\left( \nabla_\mu \phi \nabla_\nu \phi -\frac{1}{2}g_{\mu\nu} \nabla_\rho\phi \nabla^\rho \phi \right) \nonumber\\
& -\left( \nabla_\mu \nabla_\nu - g_{\mu\nu} \Box \right)\left( \frac{\alpha}{18}R+\frac{\xi}{6}\phi^2 \right)+ \left( \frac{\alpha}{72}R^2  + \frac{\lambda}{8}\phi^4 \right) g_{\mu\nu}= 0,
\end{align}
where $G_{\mu\nu}$ is the Einstein tensor, while varying the action with respect to the scalar field yields
\begin{equation}\label{scalar equation}
\Box \phi+ \left( \frac{\xi}{3}R-\lambda\phi^2 \right)\phi = 0.
\end{equation}
If we restrict ourselves to solutions with constant $\phi$ and $R$, it is trivial to see that (\ref{metric equation}) reduce to the equations of General Relativity with a cosmological constant. From (\ref{scalar equation}) is instead manifest that in order to have $R\le 0$, the scalar field has to be zero, while for $R>0$ it is possible to have both a zero and a non-zero scalar field. Therefore it will be possible to have two distinct de Sitter (or asymptotically de Sitter) solutions, and we will focus our attention to this case.

 \subsection{Einstein frame formulation}

 To simplify the discussion when dealing with a perturbed system, we first write \eqref{cosmoaction} in the Einstein frame. This is achieved by introducing the auxiliary field $\cphi$ and an auxiliary variable $\chi$ defined as
\begin{equation}
    \chi =\frac{\alpha \cphi}{18}+\frac{\xi \phi^2}{6}\ .
\end{equation}
Then, the Lagrangian in eq.\ \eqref{cosmoaction} becomes
\begin{equation}
\label{eq:auxiliaryL}
    \frac{\mathcal{L}}{\sqrt{-g}}=\chi R-\frac{\alpha \cphi^2}{36}-\frac{1}{2}\partial_{\mu}\phi\partial^{\mu}\phi-\frac{\lambda}{4}\phi^4\ .
\end{equation}
The equation of motion for $\cphi$ is simply $\cphi =R$, thus, in terms of the field variable $\chi$, the above Lagrangian can be recast in the form
\begin{equation}
    \frac{\mathcal{L}}{\sqrt{-g}}=\chi R-\frac{1}{2}(\partial \phi)^2+\frac{3 \xi}{\alpha}\phi^2 \chi -\left(\frac{\lambda}{4}+\frac{\xi^2}{4\alpha}\right)\phi^4-\frac{9}{\alpha}\chi^2 \,.
\end{equation}
Now let $\omega=M\sqrt{3/2}\ln(2\xi/M^2)$. Upon the conformal transformation $g_{\mu\nu}\rightarrow \exp(2\omega/\sqrt{6}M)g_{\mu\nu} $, one can write the Lagrangian in the Einstein frame, where it reads 
\bea
\label{eq:finalL}
\frac{\mathcal{L}}{\sqrt{-g}}= {M^{2}\over 2}R-{3M^{2}\over \cf^{2}}(\partial \cf)^{2}-{\cf^{2}\over 2M^{2}}(\partial\phi)^{2}+{3\xi \phi^{2}\cf^{2}\over 2\alpha} -{\Omega\phi^{4}\cf^{4}\over 4\alpha  M^{4}} -{9M^{4}\over 4\alpha}\ ,
\eea
and where we defined $\cf=M\,e^{-{{\omega}\over \sqrt{6}M}}$. Here, $M$ is an arbitrary parameter with mass dimensions\footnote{Note that, despite the appearance of a mass parameter, the action is still scale-invariant. In fact, $M$ turns out to be a so-called redundant parameter, see discussion in \cite{Rinaldi:2015uvu}.}.
Finally, upon the scalar fields redefinition
\begin{align}
    \zeta &= \sqrt{6} M \text{arcsinh}\left(\frac{\mathfrak{f}\phi}{\sqrt{6}M^2}\right),\\
    \rho &= \frac{M}{2} \ln\left( \frac{\phi^2}{2M^2} + \frac{3M^2}{\mathfrak{f}^2} \right),
\end{align}
we obtain the expression for the Lagrangian that we will use in the following (for full details see \cite{Ghoshal:2022qxk}), namely
\begin{equation}
\label{Einstein frame Lagrangian}
    \frac{\mathcal{L}}{\sqrt{-g}} =  \frac{M^2}{2}R - \frac{1}{2} (\partial\zeta)^2 -3\cosh^2\left( \frac{\zeta}{\sqrt{6}M} \right)(\partial \rho)^2 -U(\zeta)\,,
\end{equation}
where 
\bea\label{poten}
U(\zeta)=-{9\xi M^{4}\over \alpha}\sinh^2\left(\zeta\over \sqrt{6}M\right)+{9\Omega M^{4}\over \alpha}\sinh^4\left(\zeta\over \sqrt{6}M\right)+{9M^{4}\over 4\alpha}\,.
\eea
This representation, in the Einstein frame, of the original action is particularly convenient since the potential now depends on one field only. The field equations for the metric and scalar fields are
\begin{subequations}\label{field equations}
\begin{align}
    M^2 G_{\mu\nu} &= 6 \cosh^2{\left(\frac{\zeta}{\sqrt{6}M}\right)} \left( \partial_\mu \rho \partial_\nu \rho -\frac{1}{2} g_{\mu\nu} \partial_\lambda \rho \partial^\lambda \rho\right)\nonumber \\
   & + \partial_\mu \zeta \partial_\nu \zeta -\frac{1}{2} g_{\mu\nu} \partial_\lambda \zeta \partial^\lambda \zeta - g_{\mu\nu} U(\zeta),\\
   \Box \zeta & =  \frac{\sqrt{6}}{M} \cosh \left( \frac{\zeta}{\sqrt{6}M} \right) \sinh \left( \frac{\zeta}{\sqrt{6}M}\right) \partial_\mu \rho \partial^\mu \rho + \frac{dU}{d\zeta},\\
    \Box \rho &=  -\frac{2}{\sqrt{6}M} \tanh \left( \frac{\zeta}{\sqrt{6}M}\right) \partial_\mu \rho \partial^\mu \rho.
\end{align}
\end{subequations}
Note that $\rho=$ const (not necessarily vanishing) is a trivial solution of the system precisely because the potential depends on $\zeta$ only.

\subsection{No-hair-like theorem for scale invariant gravity}\label{subsec: theorem}

As can be seen manifestly from the Einstein frame formulation, scale invariant gravity has two additional degrees of freedom with respect to General Relativity. In this section we will show that in many relevant cases, as the ones that we will consider, these degrees of freedom have to satisfy a constraint, and the deviations from General Relativity can be characterized by a single scalar field. Thanks to this theorem, to study a Schwarzschild, de Sitter or Schwarzschild-(anti) de Sitter solution it is not necessary to impose a constant scalar curvature \textit{and} constant scalar field separately, as one follows to the other.
The trace of the metric equations of motion in the Jordan frame can be written as
\begin{equation}\label{trace}
\nabla^\mu\nabla_\mu\left(R+\frac{3(1+2\xi)}{2\alpha}\phi^2\right)=0,
\end{equation}
where the quantity in brackets is directly linked to the ``spectator'' field $\rho$ of the Einstein frame formulation as
\begin{equation}\label{conseinstein}
R+\frac{3(1+2\xi)}{2\alpha}\phi^2=\frac{3M^2}{\alpha}\mathrm{e}^{2\rho/M}
\end{equation}
whenever the equations of motion for the auxiliary field are satisfied. The distinctive form of equation (\ref{trace}) opens the possibility of formulating no-hair-like theorems for symmetric spacetimes. Let us consider a metric in the form
\begin{equation}\label{symmet}
\mathrm{d}s^2=\tau(x)^2\sigma_{ab}(y)\mathrm{d}y^a\mathrm{d}y^b+\xi_{ij}(x)\mathrm{d}x^i\mathrm{d}x^j,
\end{equation}
where $\sigma_{ab}(y)$ is the metric of a maximally symmetric manifold with dimension $d_y<4$ and $\xi_{ij}(x)$ is the metric of a generic manifold of dimension $4-d_y$. If we impose that the scalar field has the same symmetries of the spacetime, i.e. that it depends only on the $x$ variables, equation (\ref{trace}) becomes
\begin{equation}
D^iD_i\left(R+\frac{3(1+2\xi)}{2\alpha}\phi^2\right)+\frac{d_y}{\tau}D^i\tau D_i\left(R+\frac{3(1+2\xi)}{2\alpha}\phi^2\right)=0,
\end{equation}
where the $D_i$ are the covariant derivatives defined by the metric $\xi_{ij}(x)$. Multiplying by $\tau^{d_y}\left(R+\frac{3(1+2\xi)}{2\alpha}\phi^2\right)$ and integrating over a submanifold $\Sigma$ defined by the $x$ coordinates we get
\begin{equation}
\begin{split}
\int_\Sigma\mathrm{d}^{4-d_y}x\sqrt{|\xi|}\Bigg[&\tau^{d_y}\left(R+\frac{3(1+2\xi)}{2\alpha}\phi^2\right)D^iD_i\left(R+\frac{3(1+2\xi)}{2\alpha}\phi^2\right)\\
&+d_y\tau^{d_y-1}\left(R+\frac{3(1+2\xi)}{2\alpha}\phi^2\right)D^i\tau D_i\left(R+\frac{3(1+2\xi)}{2\alpha}\phi^2\right)\Bigg]=0,
\end{split}
\end{equation}
that, integrated by parts, leads to 
\begin{equation}
\begin{split}
&\left[\tau^{d_y}\left(R+\frac{3(1+2\xi)}{2\alpha}\phi^2\right)D^i\left(R+\frac{3(1+2\xi)}{2\alpha}\phi^2\right)\right]_{\partial\Sigma}\\
&-\int_\Sigma\mathrm{d}^{4-d_y}x\sqrt{|\xi|}\left[\tau^{d_y} D^i\left(R+\frac{3(1+2\xi)}{2\alpha}\phi^2\right) D_i\left(R+\frac{3(1+2\xi)}{2\alpha}\phi^2\right)\right]=0.
\end{split}
\end{equation}
If the metric $\xi_{ij}(x)$ is either positive or negative definite in $\Sigma$ and the boundary term is vanishing, then the combination $R+\frac{3(1+2\xi)}{2\alpha}\phi^2$ is constant in all $\Sigma$. Thanks to (\ref{conseinstein}) is then clear that in the Einstein frame formulation the $\rho$ field is not only a spectator field, but in many cases is also trivial.\\
This general statement becomes meaningful when specific cases are considered: with a static metric
\begin{equation}
ds^2=-\tau^2(x)\mathrm{d}t^2+\xi_{ij}(x)\mathrm{d}x^i\mathrm{d}x^j,
\end{equation}
the boundary term is vanishing whenever the boundary is taken on event horizons $\left(\tau^2=0\right)$, or on hypersurfaces of constant scalar curvature and scalar field, and it leads to 
\begin{equation}\label{conspace}
R+\frac{3(1+2\xi)}{2\alpha}\phi^2=const.\ \left(\text{in space}\right);
\end{equation}
with a Friedmann-Lematre-Robertson-Walker (FLRW) metric
\begin{equation}
ds^2=-\xi(t)\mathrm{d}t^2+\tau(t)^2\sigma_{ab}(y)\mathrm{d}y^a\mathrm{d}y^b,
\end{equation}
the boundary term is vanishing whenever the boundary is taken on cosmological singularities $\left(\tau^2=0\right)$, or on times at which the scalar curvature and scalar field are constant, and it leads to
\begin{equation}\label{conserved}
R+\frac{3(1+2\xi)}{2\alpha}\phi^2=const.\ \left(\text{in time}\right).
\end{equation}
The spatial constraint (\ref{conspace}) is then ensured for the external region of black holes which are either asymptotically flat or asymptotically anti de Sitter, and in the region between the black hole and cosmological horizon whenever both are present; these are precisely the regions that form the Euclidean sector of the Scharzschild and Schwarzschild-(anti) de Sitter metrics that we will study in section \ref{sec: thermo stability}. The time constraint (\ref{conserved}) is instead ensured if the cosmological evolution is between two fixed points, and can be used to keep track of the value of the effective cosmological constant throughout a cosmological evolution.

\section{Solutions of scale-invariant gravity}\label{sec: solutions}

\subsection{Quadratic scale-invariant cosmology }\label{subsec: cosmology}

Before considering black holes, we briefly summarize the results of \cite{Rinaldi:2015uvu,Ghoshal:2022qxk}. By imposing the flat Robertson-Walker metric $ds^2=-dt^2+a(t)^2\delta_{ij}dx^idx^j$ in the equations of motion, one finds two second-order equations in the scalar field $\phi$ and the Hubble parameter $H=\dot a/a$, namely
 \bea\label{eomt}
&&\ddot\phi+3H\dot \phi-2\xi\phi\dot H-\phi(4\xi H^{2}-\lambda\phi^{2})=0\,,\\\non
&& \alpha\left(2H\ddot H  -\dot H^{2}+6H^{2}\dot H  \right)-{1\over 2}\dot\phi^{2}+2\xi\phi \dot \phi H+{\phi^{2}\over 4}(4\xi H^{2}-\lambda \phi^{2})=0\,.
\eea

Despite the apparent complicate form, these equations can be reduced to 4 first-order non linear equations and studied with the means of dynamical system analysis. The result is that the system has only two fixed points. The first, at
\begin{equation}
H_u=\sqrt{\frac{\Lambda_u}{3}},\qquad\qquad\phi_u=0,
\end{equation}
where $\Lambda_u$ is an arbitrary cosmological constant, is unstable, while the second, at
\begin{equation}
H_s=\sqrt{\frac{\Lambda_s}{3}},\qquad\qquad\phi_s=2\sqrt{\frac{\xi\Lambda_s}{3\lambda}},
\end{equation}
where $\Lambda_s$ is still arbitrary, is an attractor. It turns out that the evolution from the unstable to the stable point yields an inflationary phase, thoroughly studied in \cite{Ghoshal:2022qxk}. The interesting feature is that, at the stable fixed point, the scalar field has a non-trivial value and the coefficient of $R$ in the action \eqref{cosmoaction} classically stabilizes around at a certain value that can be identified with the Planck mass squared, while the other terms become negligible. As a result, a mass term is spontaneously generated and the scale-invariance is broken. In particular we can use (\ref{conserved}) to relate the two cosmological constants of the two possible de Sitter solutions with zero and non-zero scalar field:
\begin{equation}\label{coscons}
\Lambda_u=\left(1+\frac{\xi\left(1+2\xi\right)}{2\alpha\lambda}\right)\Lambda_s.
\end{equation}
One can then wonder if a similar situation occurs in the space of static or stationary and spherically symmetric solutions. In the next section we show that there are two solutions corresponding to two distinct asymptotically de Sitter black holes.

\subsection{Quadratic scale-invariant black holes }\label{subsec: black holes}

The two cosmological solutions discussed in \cite{Rinaldi:2015uvu} have a static, spherically symmetric counterpart in terms of Schwarzschild-de Sitter spacetimes (SdS), with line element
\begin{equation}\label{metric}
 ds^2 = -f(r) dt^2 + \frac{1}{f(r)} dr^2 + r^2 (d\theta^2 + \sin^2{\theta} d\varphi^2),
\end{equation}
where
\begin{equation}
    f(r) = 1 -\frac{2m}{r} - \frac{\Lambda}{3} r^2.
\end{equation}
Note that both $m$ and $\Lambda$ are arbitrary integration constants, while in standard GR the cosmological constant is fixed at the level of the action. When they satisfy the relation
\begin{equation}
9m^2\Lambda<1,
\end{equation}
then the function $f(r)$ has three real roots given by
\begin{align}\label{radii}
r_b &= \frac{2}{\sqrt{\Lambda}} \cos{\left(\frac{\pi+\eta}{3}\right)} ,\\
r_c &=\frac{2}{\sqrt{\Lambda}} \cos{\left(\frac{\pi-\eta }{3}\right)},\\
r_0 &= -\frac{2}{\sqrt{\Lambda}} \cos{\left(\frac{\eta}{3}\right)},
\end{align}
where $\cos{\eta} = 3 m \sqrt{\Lambda}$. The first two are positive  and they represent the black hole and cosmological horizons, respectively.
As in the cosmological case, the two solutions are characterized by constant configurations for the scalar field, as one can easily verify by substituting the metric \eqref{metric} into the equations of motion.
Although the very purpose of this work is the study of the stability of such solutions, from now on  we will refer to them as unstable and stable solution, in analogy with \cite{Rinaldi:2015uvu}. The unstable solution is characterized by
\begin{equation}
R_u=4\Lambda_u,\qquad\qquad\phi_u = 0,
\end{equation}
while the stable one is obtained setting
\begin{equation}
R_s=4\Lambda_s,\qquad\qquad\phi_s = 2\sqrt{\frac{\xi\Lambda_s}{3\lambda}},
\end{equation}
in perfect agreement with the cosmological case. In the Einstein frame, the two SdS solutions presented above are still characterized by constant scalar fields whose values are mapped to the new variables as
\begin{subequations}\label{unstable solution Einstein frame}
    \begin{align}
    \rho_u &=\text{arbitrary},\\
    \zeta_u &=0,\\
    \Lambda_u &= \frac{9M^2}{4\alpha},\label{Lambda unstable}
\end{align}
\end{subequations}
for the unstable solution, and
\begin{subequations}\label{stable solution Einstein frame}
\begin{align}
    \rho_s &=\text{arbitrary},\\
    \zeta_s &=\sqrt{6} M \text{arcsin}\left(\sqrt{\frac{\xi}{2(\alpha \lambda + \xi^2)}}
    \right),\\
    \Lambda_s &= \frac{9M^2 \lambda}{4(\alpha \lambda + \xi^2)},\label{Lambda stable}
\end{align}
\end{subequations}
for the stable one. Note that in both cases, even if we have to fix its expression, the cosmological constant is still arbitrary because of its dependence on $M$.

\section{Black hole perturbations}\label{sec: dyn stability}

\begin{figure}[t]
  \centering
  \includegraphics[width=1\linewidth]{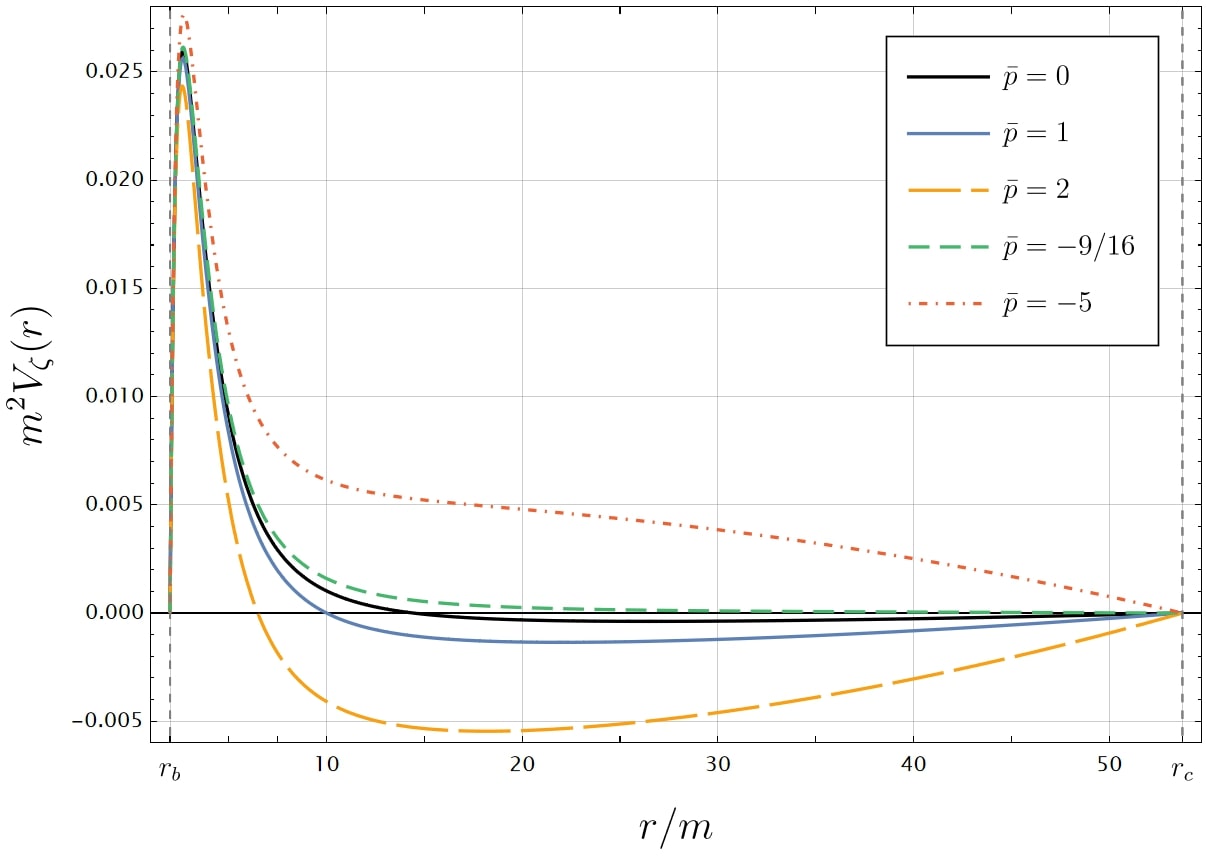}
\caption{Effective potential within the static region $r_b<r<r_c$ for $\ell=0$ and different values of $\bar{p}$. The cosmological constant is set to $\Lambda=0.001/m^2$. The $\bar{p}=0$ curve represents the GR case.}
\label{fig: potential}
\end{figure}

\begin{figure}[t]
  \centering
  \includegraphics[width=1\linewidth]{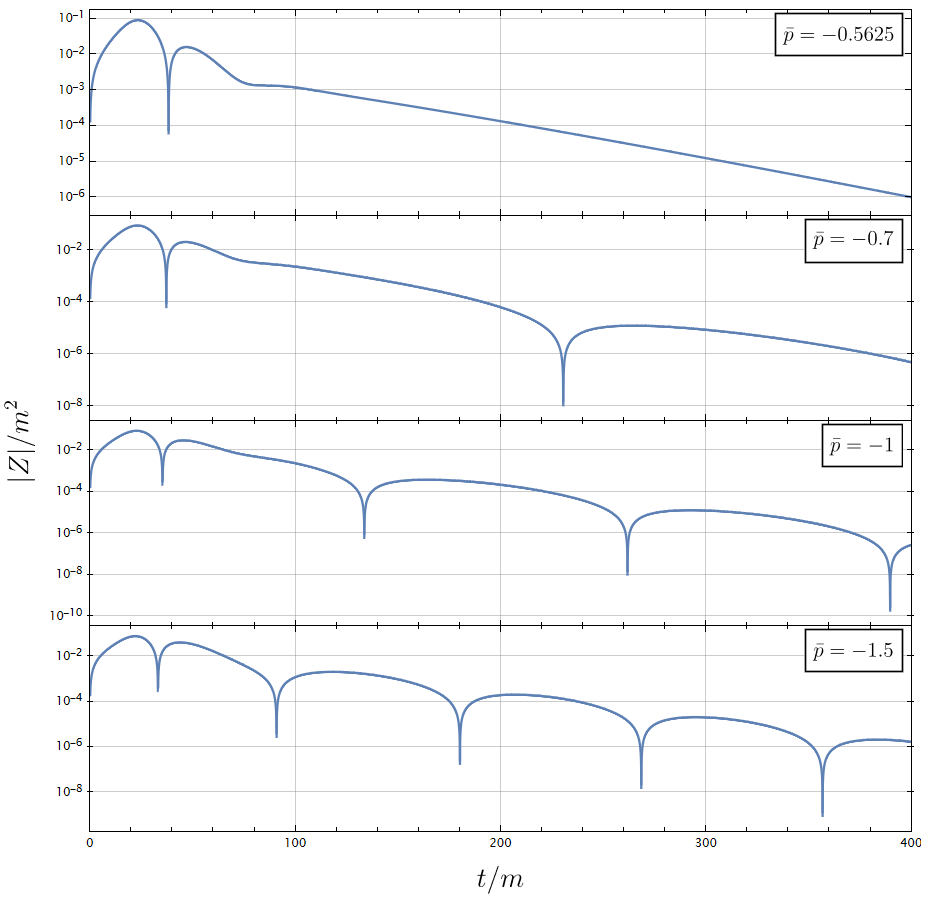}
\caption{Evolution of $|Z|/m^2$ as a function of $t/m$ at fixed radius, for $\ell=0$ and different values of $\bar{p}$, showing the onset of the oscillatory regime in the late-time region for $\bar{p}<-9/16=-0.5625$. The cosmological constant is set to $\Lambda=0.001/m^2$.}
\label{fig: oscillations}
\end{figure}

To study the dynamical stability of the two SdS solutions, it is convenient to work in the Einstein frame, where the Lagrangian density is given by \eqref{Einstein frame Lagrangian}. We now proceed to study the dynamics of metric and scalar perturbations around the two solutions \eqref{unstable solution Einstein frame} and \eqref{stable solution Einstein frame}. The fields can be expanded to linear order as
\begin{align}
    g_{\mu\nu} &= \bar{g}_{\mu\nu} + h_{\mu\nu},\\
    \rho &= \bar{\rho} + \delta\rho,\\
    \zeta &= \bar{\zeta} + \delta\zeta,
\end{align}
where barred quantities represent either the stable or unstable background solution and we will also use $\bar{\Lambda}$ to denote \eqref{Lambda unstable} or \eqref{Lambda stable}, indifferently. At first order, the system \eqref{field equations} gives
\begin{align}
M^2 \delta G_{\mu\nu} + U(\bar{\zeta}) h_{\mu\nu} &= 0,\\
\bar{\Box} \delta \rho &= 0, \label{eq delta rho}\\
\left(\bar{\Box} - \frac{d^2U}{d\zeta^2}
(\bar{\zeta}) \right) \delta \zeta&= 0,\label{eq delta zeta}
\end{align}
where $\delta G_{\mu\nu}$ is the first order perturbation of the Einstein tensor and $\bar{\Box}$ denotes the d'Alambert operator built with the background metric. We see that, at linear order, the three perturbations are decoupled. Moreover, since $U(\bar{\zeta}) = M^2 \bar{\Lambda}$, the first equation coincides with the GR equation for the metric perturbation on a SdS background with cosmological constant $\bar{\Lambda}$. Hence, the dynamics of $h_{\mu\nu}$ is not modified. The same happens for \eqref{eq delta rho}, which is nothing but the usual Klein-Gordon equation on a SdS background. The dynamics of $\delta\zeta$ is instead modified by the presence of the mass term in \eqref{eq delta zeta}. Therefore, in the following we will only focus on the dynamics of $\delta\zeta$, since instabilities can only arise from the latter. To proceed, one can exploit the spherical symmetry of the problem, adopting the following harmonic expansion for the scalar perturbation:
\begin{equation}
    \delta\zeta = \frac{Z(r)}{r} Y^{\ell \ell'}(\theta,\varphi) e^{-i\omega t},
\end{equation}
where $Y^{\ell \ell'}$ are the standard spherical harmonics. Substituting into \eqref{eq delta zeta}, one finds
\begin{equation}
    \frac{d^2Z}{dr_*^2} + \left[ \omega^2 - V_\zeta(r) \right] Z = 0,\label{eq Z}
\end{equation}
where $r_*$ is the tortoise coordinate defined by
\begin{equation}\label{tortoise coord}
\frac{dr^*}{dr} = \frac{1}{f(r)},
\end{equation}
and the argument of the potential must be considered as a function of $r^*$, inverting its definition in terms of $r$.
The effective potential featuring \eqref{eq Z} can be written in terms of a parameter $\bar{p}$ as
\begin{equation}\label{effective potential}
    V_\zeta (r) = V_{GR}(r) - \frac{4\bar{\Lambda}}{3}\bar{p}f(r),
\end{equation}
where $V_{GR}(r)$ represent the effective potential appearing in the GR case, namely for a Klein-Gordon equation on a SdS background:
 \begin{equation}
   V_{GR} (r) = f(r) \left( \frac{\ell(\ell+1)}{r^2} + \frac{2m}{r^3} -\frac{2\bar{\Lambda}}{3} \right).
\end{equation}
For $\bar{p}=0$ one recovers the GR case, while deviations for non-vanishing $\bar{p}$ can be appreciated in figure \ref{fig: potential}.
The value of the parameter $\bar{p}$ depends on the specific solution one is considering. For the unstable solution we have
\begin{equation}
    \bar{p}=p_u \equiv \xi,
\end{equation}
while the stable one is characterized by
\begin{equation}\label{P stable}
    \bar{p}=p_s \equiv 1 - \frac{9 M^2}{4\alpha\Lambda_s}(1+2\xi) = - \frac{\xi(2\alpha\lambda+\xi+2\xi^2)}{\alpha\lambda},
\end{equation}
where in the last step we used \eqref{Lambda stable}.
We will now show that the asymptotic behaviour of the perturbation crucially depends on the value of $\bar{p}$. The critical value of $\bar{p}$ separating stable from unstable modes is related to the form of the potential in the pure de Sitter, $m \rightarrow 0$ limit. This was shown in \cite{PhysRevD.60.064003}, where the case of a non-minimally coupled scalar field on a SdS background is considered. It turns out that the effective potential \eqref{effective potential} is very similar to the one computed in \cite{PhysRevD.60.064003}. In that case, starting from
\begin{equation}
    (\Box - \nu R)\phi=0,
\end{equation}
one gets
\begin{equation}
    V(r) = f(r) \left( \frac{\ell(\ell+1)}{r^2} + \frac{2m}{r^3} -\frac{2\Lambda}{3}(1-6\nu) \right).
\end{equation}
Therefore, the two potentials are mapped into each other via the substitution $\nu \leftrightarrow -\bar{p}/3$, which allows to obtain the late time behavior of the scalar perturbation from the results of \cite{PhysRevD.60.064003} as
\begin{equation}
 \delta\zeta \sim e^{-\mu \kappa_c t},
\end{equation}
where $\kappa_c$ is the surface gravity of the cosmological horizon, i.e.
\begin{equation}
\kappa_c = \frac{\Lambda (r_c - r_b)(r_c-r_0)}{6r_c},
\end{equation}
and
\begin{equation}
   \mu = \ell + \frac{3}{2} - \frac{1}{2} \sqrt{9+16\bar{p}} + O\left( \frac{r_b}{r_c} \right).
\end{equation}
We can identify three different behaviours, for different values of the parameter $\bar{p}$. When $\bar{p}<-9/16$, $\mu$ becomes complex and the purely exponential decay is replaced by a damped oscillatory regime, as observed in \cite{PhysRevD.60.064003} (see fig. \ref{fig: oscillations}):
\begin{equation}
    \delta\zeta \sim e^{-(\ell+\frac{3}{2})\kappa_c t} e^{\frac{i}{2}\sqrt{\abs{9+16\bar{p}} 
    }\kappa_c t}.
\end{equation}
The main difference with respect to \cite{PhysRevD.60.064003}, is that now $\mu$ changes sign for $\bar{p}$ greater than the critical value
\begin{equation}
    \bar{p}_c = \frac{(2\ell+3)^2-9}{16},
\end{equation}
so that we have an exponential decay for $-9/16<\bar{p}<\bar{p}_c$, with no oscillations, while the case $\bar{p}>\bar{p}_c$ yields unstable diverging modes.

\begin{figure}[t]
  \centering
  \includegraphics[width=1\linewidth]{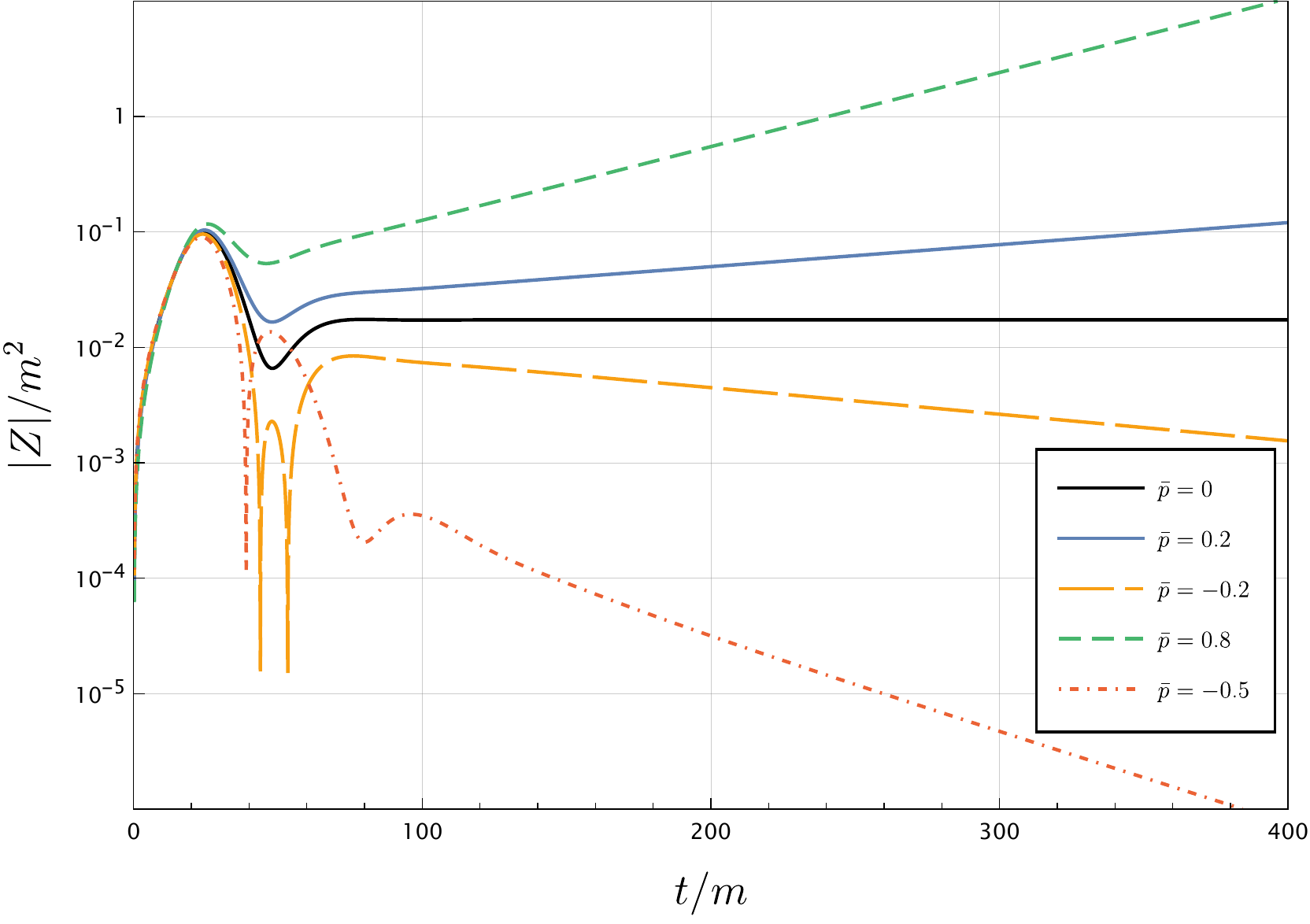}
\caption{Evolution of $|Z|/m^2$ as a function of $t/m$ at fixed radius, for $\ell=0$ and different values of $\bar{p}$. The cosmological constant is set to $\Lambda=0.001/m^2$.}
\label{fig: stability}
\end{figure}

This is confirmed by numerical integrations. To solve \eqref{eq Z} numerically, we follow the well-established method of \cite{PhysRevD.49.883,PhysRevD.63.084001,PhysRevD.70.064025,Boudet_2022}. First, we can introduce the light-cone variables $u=t-r_*$ and $v=t+r_*$, in terms of which the equation reads
\begin{equation}
     4\frac{\partial^2 Z}{\partial u \partial v} + V_{\zeta}(r) Z = 0,
\end{equation}
Then, the numerical integration is based on a discretization of the $u-v$ plane by a lattice spacing $\Delta$ where $(u,v)$ range between zero and $(u_{max},v_{max})$. The discretized version of the equation is
\begin{equation}\label{eq Z discrete}
    Z_N = Z_W + Z_E - Z_S + \frac{\Delta^2}{2} V_\zeta(r_c) \left(Z_W + Z_E \right),
\end{equation}
where a subscript means that the function is evaluated at the points $S=(u,v)$, $W=(u+\Delta,v)$, $E=(u,v+\Delta)$, $N=(u+\Delta,v+\Delta)$, while the potential is computed at $r_c=(u+\Delta/2,v+\Delta/2)$. Beside a first transient regime, the evolution of the perturbation and especially its late-time asymptotic behaviour are insensible to the initial conditions. In the results presented in the following, we assigned boundary conditions on the two axes $u=0$ and $v=0$, following \cite{PhysRevD.81.124021}. In particular, we set $Z(u,0)=Z(u,0)=0$, and Gaussian initial data on the $v$ axis, i.e. $Z(0,v)= \exp[-(v-v_c)^2/2\sigma]$. The integration starts by computing the value of $Z_N$ via \eqref{eq Z discrete}, using the initial conditions at the three points $S$, $W$, and $E$. Then, one proceeds increasing $v$ and eventually completing the grid row by row.
In order to compute the potential, at every step of the integration the tortoise coordinate must be inverted obtaining $r$ as a function of $r^*$. We do this numerically solving the explicit definition of the latter, that starting from \eqref{tortoise coord} can be written as
\begin{equation}
    r^* = \ln{\left( \frac{(r-r_b)^{\beta_b}(r+r_c+r_b)^{\beta_0}}{(r_c-r)^{\beta_c}} \right)},
\end{equation}
where
\begin{align}
    \beta_c &= \frac{3r_c}{\Lambda (r_c-r_b)(2r_c+r_b)},\\
    \beta_b &= \frac{3r_b}{\Lambda (r_c-r_b)(2r_b+r_c)},\\
    \beta_0 &= \frac{3(r_b+r_c)}{\Lambda (2r_c+r_b)(2r_b+r_c)}.
\end{align}
At the end of the integration, the perturbation evaluated at some constant $r^*$ as a function of time  can be extracted as $Z(t)=Z(t-r_*,t+r_*)$. The results presented here are obtained setting: $u_{max}=v_{max}=500$, $\Delta=0.1$, $v_c=10$, $\sigma=1$ and the function is extracted at $r_*=50m$. We work with adimensional variables rescaling every quantity by the appropriate power of the black hole mass $m$. The outcome of the numerical integrations is shown in figure \ref{fig: oscillations} and \ref{fig: stability}. In particular, figure \ref{fig: oscillations} shows the transition from an exponential decay to the damped oscillatory regime when $\bar{p}$ becomes lower than $-9/16$. Figure \ref{fig: stability} instead, confirms the stability behaviour discussed above, showing how the perturbations diverge if $\bar{p}>\bar{p}_c$, while they are stable and decay exponentially to zero when $\bar{p}<\bar{p}_c$.
Let us now apply these results to the two solutions under consideration. Given that $\bar{p}_c$ is non negative, instabilities can never trigger when the stable solution is considered, since the latter has $\bar{p}=p_s<0$ (see equation \eqref{P stable}).
Regarding the other solution instead, since $p_u=\xi$ the stability condition is provided by
\begin{equation}\label{stability condition}
\xi<\bar{p}_c.    
\end{equation}
The value of $\bar{p}_c$ is increasing with the angular momentum number $l$, implying that the first multiple suffering from instabilities would be the lowest lying mode, namely the monopole, identified by $\ell=0$. Since $\bar{p}_c(\ell=0)=0$ and being $\xi$ non negative, we conclude that the stability condition \eqref{stability condition} is never satisfied and the second solution is always unstable. Therefore, the stability character of the two solutions agrees with the one resulting from the investigations pursued in cosmological setting in \cite{Rinaldi:2015uvu}.

\section{Euclidean action and non-linear stability}\label{sec: thermo stability}

The linear analysis of the previous section has rigorously shown that the Schwarzschild-de Sitter solution with zero scalar field is always unstable, while the one with a non-zero scalar field is always stable. What cannot be inferred by the linear analysis, however, is whether one solution can have a transition into the other or not, and what is the role of black holes in the cosmological evolution shown in subsection \ref{subsec: cosmology}. A non-linear analysis can give an answer to both these problem, as well as confirm that there is no strong non-linear effect that spoils the results of the linear analysis. Usually the non-linear stability of black holes is inferred from their Thermodynamics, which however is not always well defined. The standard thermodynamical description of black holes is tightly linked with the path integral formulation of quantum gravity \cite{Gibbons:1976ue,Hawking:1978jz}. The partition function
\begin{equation}\label{therpar}
Z=\displaystyle\sum_{g}\langle g|\mathrm{e}^{-\beta H}|g\rangle,
\end{equation}
can be equivalently formulated as the partition function of an Euclidean path integral
\begin{equation}\label{pathpar}
Z=\int \mathcal{D}g\,\Big[\mathrm{e}^{-\mathcal{I}(g)}\Big],
\end{equation}
with the Euclidean time having periodicity $\beta$. Calculations can be made resorting to the semiclassical approximation, which reduces the partition function to
\begin{equation}\label{pathparsem}
Z\sim\mathrm{e}^{-\mathcal{I}(g_0)},
\end{equation}
where $g_0$ is the metric which minimizes the action. Usually, the thermodynamical properties of Schwarzschild, de Sitter and Schwarzschild-anti de Sitter solutions can be extracted from (\ref{pathparsem}) \cite{Gibbons:1976ue,Hawking:1978jz,Gibbons:1977mu,Hawking:1982dh,PhysRevD.103.084034}. Unfortunately, despite the effort of many authors (see as examples \cite{Shankaranarayanan:2003ya,Robson:2019yzx}), there is still no general consensus on how to define a global temperature for the Schwarzschild-de Sitter spacetime, and a thermodynamical description of such type of solution is a much debated subject. From a thermodynamical point of view this is due to the fact that Schwarzschild-de Sitter solutions have two horizons with different temperatures, and the system is not at thermodynamical equilibrium; from a path integral point of view, it is not possible to define a time periodicity, which removes the two conical singularities emerging in the Euclidean action from the presence of two horizons. In this section therefore we will not indulge in the thermodynamical description of Schwarzschild-de Sitter black holes, but we will remain in the semiclassical description of Euclidean quantum gravity. While in a purely classical description any spacetime which is in a stationary point of the action has equivalent relevance, in the path integral formulation the absolute value of the action determines which solution has a dominant contribution to the partition function, opening the possibility of having meta-stable solutions which will tunnel away with probability
\begin{equation}\label{tunnel}
\Gamma=A\,\mathrm{e}^{-\left(\mathcal{I}_{ms}-\mathcal{I}_s\right)},
\end{equation}
where $A$ is a prefactor and $\mathcal{I}_{ms}$, $\mathcal{I}_{s}$ are the Euclidean actions of the meta-stable and stable solutions. This approach has been used to derive the Hawking-Page transition for Schwarzschild-anti de Sitter black holes \cite{Hawking:1982dh}, and we believe it will be of relevance also in our case.

In order to evaluate the Euclidean action we need to add to the action a suitable boundary term. In appendix \ref{sec: appendix a} we compute boundary terms and actions for a more general scale invariant model with all possible quadratic combinations of curvature tensors; however for that theory it is not possible to have a manageable Einstein frame, and the dynamical discussion on the stability of the solutions cannot be made. In order to have a better comparison with the classical discussion, we then analyze the euclidean version of the action (\ref{cosmoaction}), for which the boundary term is
\begin{equation}
\mathcal{I}_{\partial M}=\int\mathrm{d}^3x\sqrt{-h}\left(\frac{\alpha}{9}\,R+\frac{\xi}{3}\,\phi^2\right)K,
\end{equation}
where $h_{\mu\nu}$ is the metric of the boundary and $K$ is its extrinsic curvature. It can be proved that for Schwarzschild, de Sitter and Schwarzschild-(anti) de Sitter spacetimes the Euclidean scale-invariant action $\mathcal{I}_{SI}$ is
\begin{equation}\label{sigr}
\mathcal{I}_{SI}=\frac{4}{9}\left(\alpha+\epsilon\frac{\xi^2}{\lambda}\right)\frac{\Lambda}{M_p^2}\mathcal{I}_{GR},
\end{equation}
where $M_p$ is the reduced Planck mass, $\epsilon=0,1$ respectively in the $\phi=0$ and $\phi\neq 0$ cases, and $\mathcal{I}_{GR}$ is the Euclidean action of General Relativity. A very similar relation holds for the more general scale invariant theory, as shown in appendix \ref{sec: appendix a}, which is a cross-check of the validity of the calculations that we performed to evaluate the actions. The particular cases are (dropping the $SI$ subscript)
\begin{subequations}
\begin{equation}\label{scase}
\mathcal{I}_{S}=0,
\end{equation}
\begin{equation}\label{dscase}
\mathcal{I}_{dS}=-\frac{32}{3}\pi^2\left(\alpha+\epsilon\frac{\xi^2}{\lambda}\right),
\end{equation}
\begin{equation}\label{sadscase}
\mathcal{I}_{SadS}=\frac{32}{27}\pi^2\alpha\frac{\Lambda r_b^2\left(3+\Lambda r_b^2\right)}{1-\Lambda r_b^2},
\end{equation}
\begin{equation}\label{actsds}
\mathcal{I}_{SdS}=-\frac{32}{9}\pi^2\left(\alpha+\epsilon\frac{\xi^2}{\lambda}\right)\Lambda\left(r_b^2+r_c^2\right),
\end{equation}
\end{subequations}
where $\Lambda$ is negative in the Schwarzschild-anti de Sitter case and positive in the Schwarzschild-de Sitter case. While the Schwarzschild \eqref{scase}, de Sitter \eqref{dscase} and Schwarzschild-anti de Sitter cases \eqref{sadscase} can be evaluated with the same techniques of General Relativity, the Schwarzschild-de Sitter case requires some care, due to the presence of two conical singularities which cannot be simultaneously removed with a suitable period of the Euclidean time. In \cite{Morvan:2022ybp} the Euclidean action for a Schwarzschild-de Sitter black hole in General Relativity has been calculated as
\begin{equation}
\mathcal{I}_{SdS,GR}=-\frac{A_b}{4G}-\frac{A_c}{4G}=-8\pi^2M_p^2\left(r_b^2+r_c^2\right),
\end{equation}
using the explicit expression for the conical singularities found in \cite{Fursaev:1995ef} in terms of Dirac delta functions. In the same paper \cite{Fursaev:1995ef} it is shown that is not possible to define quadratic curvature invariants with the same techniques, having to regularize a squared delta function. However, having no physical singularities in the Euclidean sector it is sensible to expect a finite value of the Euclidean action also in the scale invariant case. Moreover, in the calculation of the Euclidean action in \cite{Morvan:2022ybp} it is shown that the action does not depend on the periodicity of the Euclidean time, which acts as a tool to derive the action in the standard way and acquires physical meaning only through the thermodynamical interpretation. To evaluate the action we then exploited this fact to have more freedom in the bounds of integration of the Euclidean action. In particular, we considered three manifolds: a Schwarzschild-de Sitter from which we removed by hand the cosmological horizon, a Schwarzschild-de Sitter from which we removed the black hole horizon and a Schwarzschild-de Sitter from which we removed both horizons. We then assumed that the Euclidean action of Schwarzschild-de Sitter is the sum of the ones of the first two manifolds, minus the action of the third
\begin{equation}\label{summ}
\mathcal{I}_{SdS}=\mathcal{I}_{SdS\setminus\{c\}}+\mathcal{I}_{SdS\setminus\{b\}}-\mathcal{I}_{SdS\setminus\{c,b\}}.
\end{equation}
Starting from the third manifold, the Euclidean action is 
\begin{equation}\label{gauss}
\begin{split}
    \mathcal{I}_{SdS\setminus\{c,b\}}=&4\pi\beta\Bigg(\int_{r_b}^{r_c}\mathrm{d}r\,r^2\left(\frac{\alpha}{36}R^2+\frac{\xi}{6}\phi^2 R-\frac{1}{2}\partial^\mu\phi\partial_\mu\phi-\frac{\lambda}{4}\phi^4\right)+\\
    &+\sqrt{-g_{tt}}\left(\frac{\alpha}{9}\,R+\frac{\xi}{3}\,\phi^2\right)K\bigg\rvert_{r_c}+\sqrt{-g_{tt}}\left(\frac{\alpha}{9}\,R+\frac{\xi}{3}\,\phi^2\right)K\bigg\rvert_{r_b} \Bigg)=\\
    =& \frac{16}{9}\pi\left(\alpha+\epsilon\frac{\xi^2}{\lambda}\right)\Lambda\beta\Bigg(\frac{\Lambda}{3}\left(r_c^3-r_b^3\right)+\left(2 r_c-3M-\Lambda r_c^3\right)+\\
    &-\left(2 r_b-3M-\Lambda r_b^3\right)\Bigg)=\\
    =&\frac{32}{9}\pi\left(\alpha+\epsilon\frac{\xi^2}{\lambda}\right)\Lambda\beta\Bigg( r_c-\frac{\Lambda}{3} r_c^3-\left( r_b-\frac{\Lambda}{3} r_b^3\right)\Bigg)=0,
\end{split}
\end{equation}
where $\beta$ is a generic Euclidean time periodicity, and the last equivalence is given by the horizons definition. The other two integrals are then given by
\begin{equation}
\begin{split}
    \mathcal{I}_{SdS\setminus\{c\}}=&4\pi\beta_{b}\Bigg(\int_{r_b}^{r_c}\mathrm{d}r\,r^2\left(\frac{\alpha}{36}R^2+\frac{\xi}{6}\phi^2 R-\frac{1}{2}\partial^\mu\phi\partial_\mu\phi-\frac{\lambda}{4}\phi^4\right)+\\
    &+\sqrt{-g_{tt}}\left(\frac{\alpha}{9}\,R+\frac{\xi}{3}\,\phi^2\right)K\bigg\rvert_{r_c} \Bigg)=-4\pi\beta_{b}\sqrt{-g_{tt}}\left(\frac{\alpha}{9}\,R+\frac{\xi}{3}\,\phi^2\right)K\bigg\rvert_{r_b},\\
    \mathcal{I}_{SdS\setminus\{b\}}=&4\pi\beta_{c}\Bigg(\int_{r_b}^{r_c}\mathrm{d}r\,r^2\left(\frac{\alpha}{36}R^2+\frac{\xi}{6}\phi^2 R-\frac{1}{2}\partial^\mu\phi\partial_\mu\phi-\frac{\lambda}{4}\phi^4\right)+\\
    &+\sqrt{-g_{tt}}\left(\frac{\alpha}{9}\,R+\frac{\xi}{3}\,\phi^2\right)K\bigg\rvert_{r_b} \Bigg)=-4\pi\beta_{b}\sqrt{-g_{tt}}\left(\frac{\alpha}{9}\,R+\frac{\xi}{3}\,\phi^2\right)K\bigg\rvert_{r_c},
\end{split}
\end{equation}
where $\beta_b$ and $\beta_c$ are the periodicities required to remove respectively the black hole and cosmological horizon conical singularities, and the equivalences are given by the result in (\ref{gauss}). The two terms evaluated on the Schwarzschild-de Sitter metric then appear as
\begin{equation}
\begin{split}
    \mathcal{I}_{SdS\setminus\{c\}}=&-\frac{32}{9}\pi^2\left(\alpha+\epsilon\frac{\xi^2}{\lambda}\right)\Lambda r_b^2,\\
    \mathcal{I}_{SdS\setminus\{b\}}=&-\frac{32}{9}\pi^2\left(\alpha+\epsilon\frac{\xi^2}{\lambda}\right)\Lambda r_c^2,
\end{split}
\end{equation}
where we considered a strictly positive temperature for the cosmological horizon, as in the pure de Sitter case. It is then trivial to see that the sum (\ref{summ}) gives exactly (\ref{actsds}), which satisfies the same relation (\ref{sigr}) as all the other solutions considered, which have been calculated with standard techniques. Moreover, when applied to the context of General Relativity our procedure has exactly the same result of \cite{Morvan:2022ybp}. While it might seem a crude calculation, this result indicates that the Euclidean action of Schwarzschild-de Sitter is given entirely by the conical singularities, exactly as shown for General Relativity in \cite{Morvan:2022ybp}. The main difference here is that the singularity is integrated away while defining the action, and not during the volume integration, opening the possibility of using the same technique to all the actions for which the boundary terms are known.

\subsection{Stability of Schwarzschild-de Sitter black holes}\label{subsec: stability}

In order to understand whether an unstable SdS black hole undergoes a transition into a stable one, we have to understand which are the most convenient parameters to use to compare the two solutions. While in principle the Euclidean action of a Schwarzschild-de Sitter black hole is characterized by two parameters, that is its mass and cosmological constant, the scale invariant nature of the theory removes one degree of freedom by making everything dependent only from a dimensionless combination of the two parameters. With the definitions for the black hole and cosmological horizons (\ref{radii}), the expression (\ref{actsds}) becomes
\begin{equation}
\mathcal{I}_{SdS}=-\frac{64}{9}\pi^2\left(\alpha+\epsilon\frac{\xi^2}{\lambda}\right)\left(2-\cos\left(\frac{2\eta}{3}\right)\right),
\end{equation}
where $\cos\eta=3m\sqrt{\Lambda}$, and therefore the difference in the actions with zero and non-zero scalar field is
\begin{equation}\label{diffact}
\mathcal{I}_{SdS,\,u}-\mathcal{I}_{SdS,\,s}=\frac{64}{9}\pi^2\Bigg(\frac{\xi^2}{\lambda}\left(2-\cos\left(\frac{2\eta_s}{3}\right)\right)+\alpha\left(\cos\left(\frac{2\eta_u}{3}\right)-\cos\left(\frac{2\eta_s}{3}\right)\right)\Bigg),
\end{equation} 
with $\cos\eta_u=3m_u\sqrt{\Lambda_u}$ and $\cos\eta_s=3m_s\sqrt{\Lambda_s}$. The two parameters $\Lambda_u$ and $\Lambda_s$, however, are not independent one from the other. If we consider the black holes as immersed in a large, evolving universe, it is sensible to expect that the value of the effective cosmological constant will be given by the global cosmological evolution, and will not be affected by the presence of the black holes. The two parameters will then be linked by the relation (\ref{coscons}), and the two SdS black holes can be compared using relations between $\eta_u$ and $\eta_s$ which will depend only from the free parameters of the theory. We can now consider four different cases: black holes with the same horizons radii, and then with the same geometry, which satisfy
\begin{equation}
\eta_u=\eta_s;
\end{equation}
black holes with the same mass parameter $m$, that is for which only the cosmological constant undergoes an evolution, which satisfy
\begin{equation}
\eta_u=\arccos\left(\sqrt{1+\frac{\xi(1+2\xi)}{2\alpha\lambda}}\cos\eta_s\right);
\end{equation}
black holes with the same cosmological horizon temperature, which satisfy
\begin{equation}
\begin{split}
\eta_u=&\pi-3\arccos\Bigg(\sqrt{1+\frac{\xi(1+2\xi)}{2\alpha\lambda}}\frac{1}{8\cos\left(\frac{\pi-\eta_s}{3}\right)}\Bigg(-3+4\cos^2\left(\frac{\pi-\eta_s}{3}\right)+\\
&+\sqrt{\frac{48}{\sqrt{1+\frac{\xi(1+2\xi)}{2\alpha\lambda}}}\cos^2\left(\frac{\pi-\eta_s}{3}\right)+\left(-3+4\cos^2\left(\frac{\pi-\eta_s}{3}\right)\right)^2}\,\Bigg)\Bigg);
\end{split}
\end{equation}
and, finally, black holes with the same black hole horizon temperature, that shares also the same absorption cross section \cite{Sini:2007zz}, which satisfy
\begin{equation}
\begin{split}
\eta_u=&3\arccos\Bigg(\sqrt{1+\frac{\xi(1+2\xi)}{2\alpha\lambda}}\frac{1}{8\cos\left(\frac{\pi+\eta_s}{3}\right)}\Bigg(-3+4\cos^2\left(\frac{\pi+\eta_s}{3}\right)+\\
&+\sqrt{\frac{48}{\sqrt{1+\frac{\xi(1+2\xi)}{2\alpha\lambda}}}\cos^2\left(\frac{\pi+\eta_s}{3}\right)+\left(-3+4\cos^2\left(\frac{\pi+\eta_s}{3}\right)\right)^2}\,\Bigg)\Bigg)-\pi.
\end{split}
\end{equation}
\begin{figure}[t]
  \centering
  \includegraphics[width=1\linewidth]{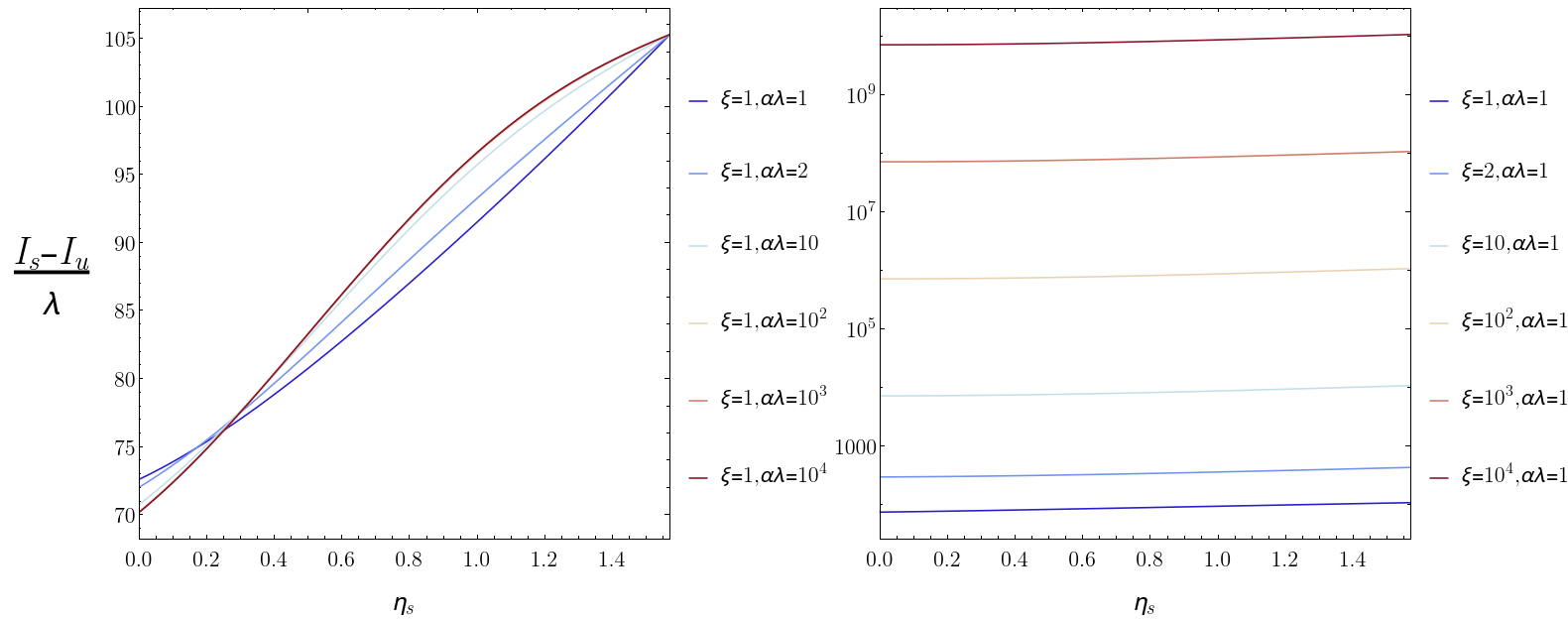}
\caption{Difference of the Euclidean actions for black holes with the same temperature at the black hole horizon, varying separately $\alpha\lambda$, in the first panel, and $\xi$ in the second panel.}
\label{fig: diffaction}
\end{figure}
While a quantitative discussion on the behaviour of (\ref{diffact}) requires fixing the three free parameters of the theory, the positiveness of the difference is assessed in all the cases under analysis simply by requiring $\alpha,\,\xi,\,\lambda>0$. If we compare black holes with the same geometry, the case is trivial because only the term multiplied by $\xi^2/\lambda$, which is always positive, survives. If we compare black holes with the same mass parameter or the same cosmological horizon temperature, we have to add the term multiplied by $\alpha$ which, however, it is also always positive taking into account that $\eta_u< \eta_s$ for $\sqrt{1+\frac{\xi(1+2\xi)}{2\alpha\lambda}}> 1$. The latter case requires some care, considering that in this case $\eta_u> \eta_s$ for $\sqrt{1+\frac{\xi(1+2\xi)}{2\alpha\lambda}}> 1$, and then the term multiplied by $\alpha$ is now negative; however as $\alpha$ is increased, the combination $1+\frac{\xi(1+2\xi)}{2\alpha\lambda}$ gets closer to 1 and the terms which is multiplied by $\alpha$ gets smaller, and there is no analytical way to assess whether the difference between the actions is positive or not. In figure \ref{fig: diffaction} we show numerically that the difference is positive both in the case of $\alpha$ being much larger than $\frac{\xi}{\lambda}$, and in the opposite case. Therefore, we can state that in all the cases under consideration
\begin{equation}
\mathcal{I}_{SdS,\,u}>\mathcal{I}_{SdS,\,s}\qquad\qquad\text{for every }m,\Lambda_u\text{ s.t. }3m\sqrt{\Lambda_u}<1,
\end{equation}
where the constraint on the parameters is given by $m<\frac{1}{3}\min\left(\frac{1}{\sqrt{\Lambda_u}},\frac{1}{\sqrt{\Lambda_s}}\right)=\frac{1}{3\sqrt{\Lambda_u}}$. Such relation guarantees that any Schwarzschild-de Sitter black hole prefers to be immersed in a de Sitter space with a smaller cosmological constant and, in our specific case, confirms that a Schwarzschild-de Sitter black hole will have a transition from a solution with zero scalar field to a solution with a non-zero scalar field.\\
\begin{figure}[t]
  \centering
  \includegraphics[width=1\linewidth]{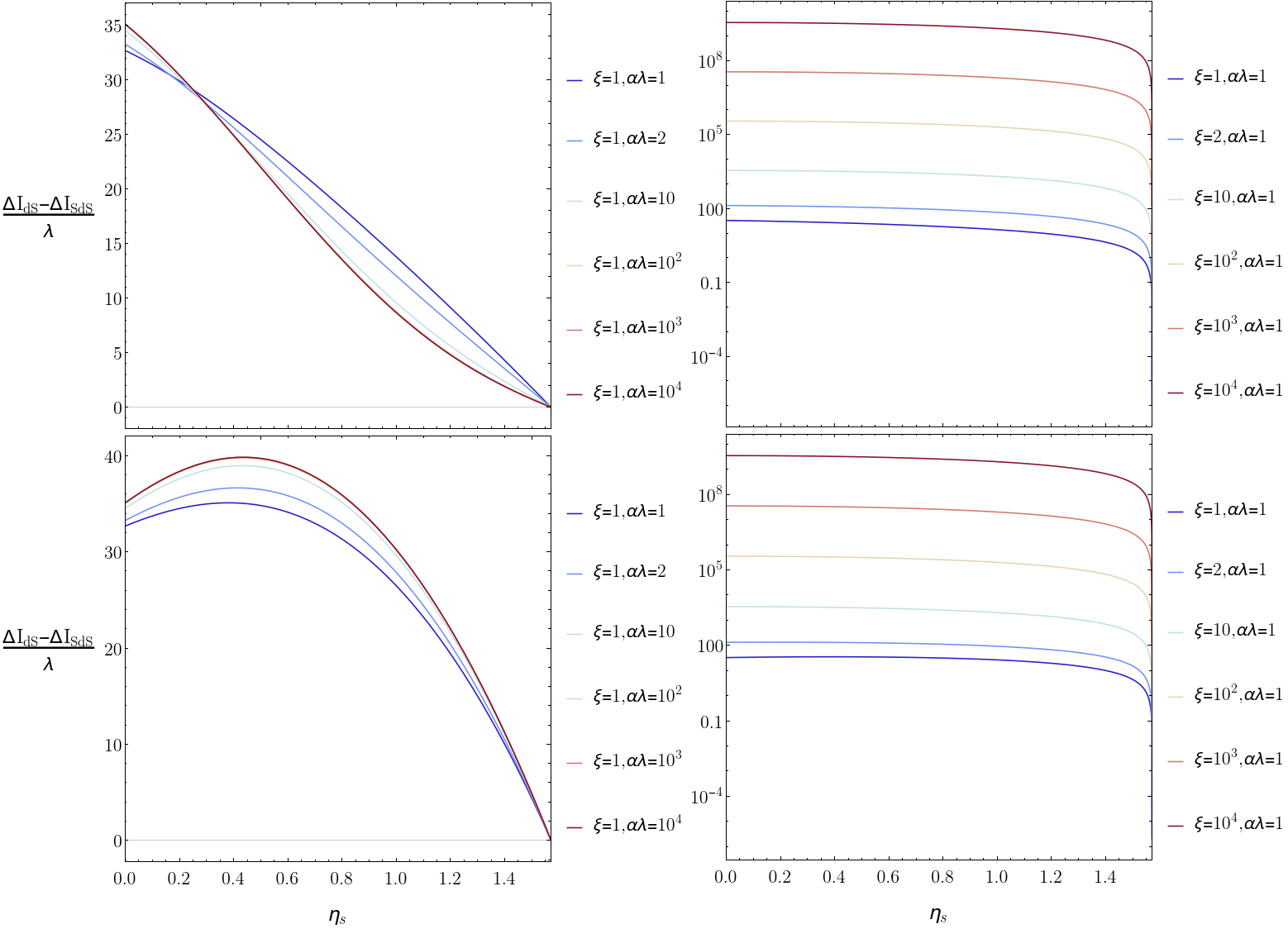}
\caption{Decrease of the tunnelling probability between two Schwarzschild-de Sitter and two de Sitter spacetimes for black holes with the same temperature at the black hole horizon, in the panels on top, and with the same temperature at the cosmological horiozn, in the panels in the bottom, varying separately $\alpha\lambda$, in the panels on the left, and $\xi$ in the panels on the right.}
\label{fig: tunnel}
\end{figure}
Having clarified that the unstable Schwarzschild-de Sitter will have a transition into the stable one, what is left to understand is if the presence of a black hole immersed in the cosmological setting described in (\ref{subsec: cosmology}) can stabilize the unstable de Sitter space, or aggravates its instability. Taking into account the difference in the Euclidean actions of the two de Sitter spacetimes
\begin{equation}\label{diffactds}
\mathcal{I}_{dS,\,u}-\mathcal{I}_{dS,\,s}=\frac{32}{3}\pi^2\frac{\xi^2}{\lambda},
\end{equation}
it is clear that, when comparing black holes by their radius, the difference (\ref{diffact}) is always smaller than (\ref{diffactds}); the tunnel probability (\ref{tunnel}) will then always be smaller if a black hole is present, and the unstable de Sitter configuration will have longer life expectation. The situation is the same for black holes with the same temperature either at the black hole or cosmological horizons, as shown in figure \ref{fig: tunnel}. Comparing black holes by their mass renders the discussion more complicated: if the parameters of the theory are such that $\alpha\lambda\ll\xi^2$, then the presence of a black hole will stabilize the de Sitter space; in the opposite case, when $\alpha\lambda\gg\xi^2$ the presence of a black hole will increase the decaying rate of the unstable de Sitter space. Intermediate values can present a situation where a small black hole will stabilize the de Sitter space, while a large one will increase its instability. As general statement, we have that the presence of a black hole will always stabilize the de Sitter space whenever the parameters of the theory satisfy
\begin{equation}
\xi\left(1+\frac{\xi^2}{\alpha\lambda}\right)>\frac{1}{2}.
\end{equation}
Taking into account the most recent results obtained in the cosmological setting \cite{Ghoshal:2022qxk}, we expect $\alpha$ to be much smaller than $\frac{\xi^2}{\lambda}$, and then the presence of black holes will always decrease the tunnelling probability of an unstable de Sitter spacetime.

\section{Conclusions}\label{sec: conclusions}

In this work we examined the static and spherically symmetric solutions of a scale-invariant tensor-scalar theory of gravity. It is known that, in such a theory, inflationary solutions exist connecting one unstable fixed point (characterized by a vanishing scalar field) to a stable one (with non-zero constant scalar field). Analogously, in the space of static and spherically symmetric solutions, we have found that there are two asymptotically de Sitter black hole configurations, corresponding to the two fixed points (one with zero and the other with non-zero scalar field). In the absence of time evolution in the system, the problem of understanding the stability of these solutions has been faced in two complementary ways. First, we studied the scalar perturbations and we showed that the solution with zero scalar field is characterized by exponentially growing perturbation modes and is actually unstable, while these modes are absent in the one with non-zero scalar field, which is actually stable. Then we analyzed the Euclidean path integral of the solutions and we found that there is a non-zero transition probability from the unstable solution to the stable one, which is however always smaller than the transition probability between two pure de Sitter solutions.

\appendix

\section{Euclidean action of extended scale-invariant theory}\label{sec: appendix a}

Here, we slightly generalize the action by including the Weyl and the Gauss Bonnet terms, that is we include all the possible quadratic invariant of the curvature tensors. In the cosmological case, these terms would be irrelevant, at least at the level of background equations of motion. 
Thus, we now consider the most general scale-invariant theory with a single scalar field, namely
\begin{equation}\label{actful}
    \mathcal{I} = \int\mathrm{d}^4x\sqrt{-g}\left[ \frac{\alpha}{36}R^2-\frac{\beta}{6}C^{\mu\nu\rho\sigma}C_{\mu\nu\rho\sigma}+\frac{\gamma}{6}\mathcal{G}+\frac{\xi}{6}\phi^2 R-\frac{1}{2}\partial^\mu\phi\partial_\mu\phi-\frac{\lambda}{4}\phi^4  \right],
\end{equation}
where $C_{\mu\nu\rho\sigma}$ is the Weyl tensor, $\mathcal{G}$ is the Gauss-Bonnet combination, and the Weyl squared term has a negative sign for consistence with the results found using the Stelle action, i.e. with $\lambda=0$ and $\phi=$ const. Setting $\beta=\gamma=0$ one recovers the quadratic scale-invariant model studied in \cite{Rinaldi:2015uvu}. We note that, thanks to the topological nature of the Gauss-Bonnet term and the traceless nature of the Weyl term, the theorems discussed in subsection \ref{subsec: theorem} are still valid.

In order to evaluate the Euclidean action we need to add to the action a suitable boundary term. Taking into account all the different terms in (\ref{actful}), the boundary terms result to be
\begin{equation}
\begin{split}
\mathcal{I}_{\partial M,\,1}=&\int\mathrm{d}^3x\sqrt{-h}\,\frac{\alpha}{9}\,R\,K,\\
\mathcal{I}_{\partial M,\,2}=&-\int\mathrm{d}^3x\sqrt{-h}\,\frac{\beta}{3}\,C^{\mu\nu\rho\sigma}\left(n_\mu n_\rho K_{\nu\sigma}-n_\mu n_\sigma K_{\nu\rho}+n_\nu n_\sigma K_{\mu\rho}-n_\nu n_\rho K_{\mu\sigma}\right),\\
\mathcal{I}_{\partial M,\,3}=&\int\mathrm{d}^3x\sqrt{-h}\,\frac{\gamma}{3}\big[ R^{\mu\nu\rho\sigma}\left(n_\mu n_\rho K_{\nu\sigma}-n_\mu n_\sigma K_{\nu\rho}+n_\nu n_\sigma K_{\mu\rho}-n_\nu n_\rho K_{\mu\sigma}\right)+\\ &-4\left(R^{\mu\nu}K_{\mu\nu}+R^{\mu\nu}n_\mu n_\nu K -2 R^{\mu\nu}n_\mu n^\rho K_{\rho\nu}\right) + 2 R\,K\big],\\
\mathcal{I}_{\partial M,\,4}=&\int\mathrm{d}^3x\sqrt{-h}\,\frac{\xi}{3}\,\phi^2\,K,
\end{split}
\end{equation}
where the ones for the Weyl and Gauss-Bonnet combinations are derived from \cite{Solodukhin:2015eca}. It can be proved that for Schwarzschild, de Sitter and Schwarzschild-(anti) de Sitter spacetimes the Euclidean scale-invariant action is
\begin{equation}\label{sigr2}
\mathcal{I}_{SI}=\frac{4}{9}\left(\alpha+\beta+\epsilon\frac{\xi^2}{\lambda}\right)\frac{\Lambda}{M_p^2}\mathcal{I}_{GR}+\frac{32}{3}\pi^2\left(\beta-\gamma\right)N_H,
\end{equation}
where $N_H$ is the number of horizons, $M_p$ is the reduced Planck mass, $\epsilon=0,1$ respectively in the $\phi=0$ and $\phi\neq 0$ cases, and $\mathcal{I}_{GR}$ is the Euclidean action of General Relativity. The particular cases are (dropping the SI pedices)
\begin{subequations}
\begin{equation}
\mathcal{I}_{S}=\frac{32}{3}\pi^2\left(\beta-\gamma\right),
\end{equation}
\begin{equation}
\mathcal{I}_{dS}=-\frac{32}{3}\pi^2\left(\alpha+\beta+\epsilon\frac{\xi^2}{\lambda}\right)+\frac{32}{3}\pi^2\left(\beta-\gamma\right),
\end{equation}
\begin{equation}
\mathcal{I}_{SadS}=\frac{32}{27}\pi^2\left(\alpha+\beta\right)\frac{\Lambda r_b^2\left(3+\Lambda r_b^2\right)}{1-\Lambda r_b^2}+\frac{32}{3}\pi^2\left(\beta-\gamma\right),
\end{equation}
\begin{equation}\label{actsds2}
\mathcal{I}_{SdS}=-\frac{32}{9}\pi^2\left(\alpha+\beta+\epsilon\frac{\xi^2}{\lambda}\right)\Lambda\left(r_b^2+r_c^2\right)+\frac{64}{3}\pi^2\left(\beta-\gamma\right).
\end{equation}
\end{subequations}
The relation (\ref{sigr2}) is simply related with (\ref{sigr}) by a redefinition of $\alpha\to\alpha+\beta$ and with the addition of a constant term $\frac{32}{3}\pi^2\left(\beta-\gamma\right)N_H$; the discussion made in subsection \ref{subsec: stability} therefore is still valid in the general theory.

\bibliographystyle{JHEP}
\providecommand{\href}[2]{#2}\begingroup\raggedright\endgroup

\end{document}